\documentclass[11pt]{article}
\usepackage{amsmath,amssymb,color,graphics,epsfig,cite}

\usepackage{amsfonts}

\newcommand{\hoch}[1]{$\, ^{#1}$}

\makeatletter
\@addtoreset{equation}{section}
\makeatother

\newcommand{\be}{\begin{equation}}
\newcommand{\ee}{\end{equation}}
\newcommand{\bea}{\setlength\arraycolsep{2pt} \begin{eqnarray}}
\newcommand{\eea}{\end{eqnarray}}
\newcommand{\nn}{\nonumber}

\def\ft#1#2{{\textstyle{\frac{\scriptstyle #1}{\scriptstyle #2} } }}
\def\fft#1#2{{\frac{#1}{#2}}}

\def\0{{\sst{(0)}}}
\def\1{{\sst{(1)}}}
\def\2{{\sst{(2)}}}
\def\3{{\sst{(3)}}}
\def\4{{\sst{(4)}}}
\def\5{{\sst{(5)}}}
\def\6{{\sst{(6)}}}
\def\7{{\sst{(7)}}}
\def\8{{\sst{(8)}}}
\def\sst#1{{\scriptscriptstyle #1}}

\def\im{{{\rm i\,}}}

\begin{document}


\begin{center}
{\Large {\bf Time Machines and AdS Solitons with Negative Mass}}

\vspace{40pt}
{\bf Xing-Hui Feng, Wei-Jian Geng\hoch{*} and  H. L\"u}

\vspace{10pt}

{\it Center for Advanced Quantum Studies, Department of Physics, \\
Beijing Normal University, Beijing 100875, China}

\vspace{40pt}

\underline{ABSTRACT}
\end{center}

We show that in $D=2n+1$ dimensions, when mass is {\it negative} and all angular momenta are non-vanishing, Kerr and Kerr-AdS metrics describe smooth time machines, with no curvature singularity.   Turning off the angular momenta appropriately can lead to static AdS solitons with negative mass.  Setting zero the cosmological constant yields a class of Ricci-flat K\"ahler metrics in $D=2n$ dimensions.  We also show that Euclidean-signatured AdS solitons with negative mass can also arise in odd dimensions.  We then construct time machines in $D=5$ minimal gauged supergravity that carry only magnetic dipole charges.  Turning off the cosmological constant, the time machine becomes massless and asymptotically  flat.  It can be described as a constant time bundle over the Eguchi-Hanson instanton.

\vfill {\footnotesize xhfengp@mail.bnu.edu.cn \ \ \ \hoch{*}gengwj@mail.bnu.edu.cn\ \ \
mrhonglu@gmail.com}

\thispagestyle{empty}

\pagebreak

\tableofcontents
\addtocontents{toc}{\protect\setcounter{tocdepth}{2}}



\section{Introduction}

This paper studies the properties of the general Kerr metrics with or without a cosmological constant, when they do not describe rotating black holes. The Kerr metric \cite{Kerr:1963ud} of a rotating black hole that is asymptotic to four-dimensional Minkowski spacetime is far more subtle to construct than the static Schwarzschild metric \cite{Schwarzschild:1916uq} with spherical symmetry. The solution was generalized by Carter \cite{Carter:1968ks} to include a cosmological constant and the metric describes a rotating back hole in de Sitter (dS) or anti-de Sitter (AdS) spacetimes for positive or negative cosmological constants respectively.  Inspired by string theory, Kerr metrics in higher dimensions were constructed in \cite{Myers:1986un}.  Kerr-(A)dS metric in five dimensions were constructed in \cite{Hawking:1998kw}, motivated by the AdS/CFT correspondence \cite{Maldacena:1997re}. The Kerr-(A)dS metrics in general dimensions were later constructed in \cite{Gibbons:2004uw,Gibbons:2004js}.

One fascinating feature of Riemannian geometry is that a local metric may extend onto very different manifolds in different coordinate patches.  For example, a five-dimensional Kerr-AdS ``over-rotating'' metric is equivalent, after performing a coordinate transformation, to an under-rotating Kerr-AdS metric \cite{Chen:2006ea}. Kerr and particularly Kerr-AdS metrics are very complicated in general dimensions and it is quite possible that these local metrics can describe spacetimes other than rotating black holes.  Indeed, we find that when the mass is negative, the local metrics in $D=2n+1$ can describe a smooth time machine, provided that all independent orthogonal angular momenta are turned on. In this paper we adopt the definition of time machine in \cite{Cvetic:2005zi}.  In such a time machine, the spacetime closes off at some Euclidean pseudo horizon at the price that the real time coordinate becomes periodic. The curvature power-law singularity is outside the spacetime. The conclusion holds for both asymptotically-flat or AdS solutions.

Turning off the angular momenta appropriately, we obtain AdS solitons with negative mass.  These solutions with general parameters are of multi-cohomogeneity.  If the starting Kerr-AdS metrics have equal angular momenta and hence are cohomogeneity one, the corresponding AdS solitons are also cohomogeneity one, with level surfaces as $S^{2n-1}/\mathbb Z_{k}$. Such a five-dimensional AdS soliton was previously constructed in \cite{Clarkson:2005qx,Clarkson:2006zk}.  Ours generalize to arbitrary $2n+1$ dimensions and multi-comohogeneity.

We can set the cosmological constants of the AdS solitons to zero, and the resulting solutions are direct products of time and a class of $D=2n$ Ricci-flat metrics.  The special case of cohomogeneity-one solutions are the Eguchi-Hanson (EH) instanton and its higher-dimensional generalizations.

The paper is organized as follows.  In section 2, we begin with the $D=5$ example, and then demonstrate that all Kerr or Kerr-AdS metrics in odd dimensions with negative mass can have the smooth time-machine configuration when all the angular momenta are turned on.  In section 3, we concentrate on Kerr-AdS metrics in odd dimensions and obtain the static limit that describes soliton configruations with negative mass.  In section 4, we turn off the cosmological constant of the soliton configurations and obtain a class of Ricci-flat metrics in $D=2n$ dimensions. In section 5, we perform Wick rotation on the Kerr metrics and find that in odd dimensions, the Euclidean-signatured solitons can also have negative mass. In section 6, we consider charged Kerr-AdS solution in five-dimensional minimal gauged supergravity and obtain the analogous limit of time machines that carry magnetic dipole charges.  Turning off the gauging, we obtain a massless asymptotically-flat time machine that is a constant time bundle over the EH instanton.  We conclude the paper in section 7.

\section{Time machines with negative mass}

In this section, we consider Kerr and Kerr-AdS metrics in odd $D=2n+1$ dimensions.  We show that when mass is negative, the metrics can describe smooth time machines where geodesic complete on some Euclidean Killing horizons, provided that all angular momenta are turned on.   The conclusion is true for both asymptotically flat or AdS metrics.  For this reason, we focus on the discussion on Kerr-AdS metrics to avoid the repetition of discussing the Kerr and Kerr-AdS metrics separately.  However, since our results are applicable for both types of metrics, we shall not emphasise the word AdS.

\subsection{$D=5$ time machines with equal angular momenta}

\subsubsection{Local metrics in $D=5$}

We start with a class of rotating metrics in five dimensions with the level surfaces as squashed $S^3$ written as a $U(1)$ bundle over $S^2$:
\bea
ds_5^2 &=& \fft{dr^2}{f} - \fft{f}{W} dt^2 + \ft14 r^2 W (\sigma_3 + \omega)^2 + \ft14 r^2 d\Omega_2^2\,,\cr
f &=& (1+g^2 r^2) W - \fft{\mu}{r^2},,\qquad W=1 + \fft{\nu}{r^4}\,,\qquad \omega = \fft{2\sqrt{\mu\nu}}{r^4W}\, dt\,.\label{d5rot1}
\eea
Here the metric $d\Omega_2^2$ and 1-form $\sigma_3$ are given by
\be
d\Omega_2^2 = d\theta^2 + \sin^2\theta\, d\phi^2\,,\qquad \sigma_3=d\psi + \cos\theta d\phi\,.
\label{s2sig3}
\ee
The metric for the unit round $S^3$ is given by
\be
d\Omega_3^2 = \ft14 \big(\sigma_3^2 + d\Omega_2^2\big)\,.
\ee
Thus the metric (\ref{d5rot1}) for constant $t$ and $r$ describes squashed $S^3$ with $W$ as the squashing parameter.

Metrics (\ref{d5rot1}) are all Einstein with $R_{\mu\nu}=-4 g^2 g_{\mu\nu}$, where constant $1/g$ is the AdS radius. The solutions are specified by two integration constants $(\mu,\nu)$.  (There should be no confusion between $(\mu,\nu)$ as the spacetime indices and as integration constants of the solutions.) The invariant Riemann tensor squared is
\be
{\rm Riem}^2=40g^4 +\fft{72(\mu-g^2\nu)^2}{r^8} - \fft{384\nu (\mu - g^2\nu)}{r^{10}} +
\fft{384\nu^2}{r^{12}}\,.
\ee
Thus there is only one power-law curvature singularity at $r=0$.  Depending on the values of the constants $(\mu,\nu)$, the metrics can extend smoothly onto very different manifolds.  When $\mu=0=\nu$, the metrics become the AdS$_5$ vacuum in global coordinates. Thus the metrics all approach AdS asymptotically at the $r\rightarrow \infty$ region. In particular, when $\mu>0$ and $\nu=0$, the metric is the well-known Schwarzschild-AdS solution.  We now give the list of $(\mu,\nu)$ values for which the power-law curvature singularity at $r=0$ can be either unreachable geodesically or hidden inside an event horizon
\begin{itemize}

\item $\mu>0$ and $\nu>0$:  Rotating black hole with equal angular momenta and positive mass, which we shall give a quick review in the next subsection \ref{sec:ad5rot}.

\item $\mu<0$ and $\nu<0$: Time machine with equal angular momenta and negative mass, which we shall discuss in \ref{sec:ads5time}.

\item $\mu=0$ and $\nu<0$:  AdS static soliton with negative mass, which we shall discuss in section \ref{sec:soliton}.

\item $\mu\nu<0$, the metric becomes real if we make a Wick rotation $t=\im \tau$, giving rise to Einstein-Riemannian geometry. We shall discuss this in section \ref{sec:instanton}.

\end{itemize}

\subsubsection{Rotating black hole}
\label{sec:ad5rot}

We first consider the case with $\mu>0$ and $\nu>0$. The metric describes a rotating black hole that is non-rotating asymptotically.  The event horizon is located at $r=r_0>0$  that is the largest real root of $f(r)$. A necessary condition for the existence of such a root is $1-\nu g^2/\mu>0$.  We can express $\mu$ in terms of $r_0$ and $\nu$:
\be
\mu=\fft{(r_0^4+\nu)(1+ g^2 r_0^2)}{r_0^2}\,.
\ee
Following the standard technique, we obtain the thermodynamical quantities including the mass $M$, angular momentum $J$, angular velocity $\Omega_+$, temperature $T$ and entropy $S$.
\bea
M&=&\ft18\pi (3\mu+ g^2\nu)\,,\qquad J=\ft14\pi \sqrt{\mu\nu}\,,\qquad \Omega_+ =\fft{2\sqrt{\nu(1 + g^2 r_0^2)}}{r_0\sqrt{r_0^4 + \nu}}\,,\cr
T &=& \fft{2g^2 r_0^6 +r_0^4 -\nu}{2\pi r_0^3 \sqrt{r_0^4 + \nu}}\,.\,,\qquad S = \ft12 \pi^2 r_0 \sqrt{r_0^4 +\nu}\,.\label{d5thermo}
\eea
These quantities satisfy the first law of black hole thermodynamics
\be
dM=T dS + \Omega_+ dJ\,.
\ee
Note that in five dimensions, there are in general two independent angular momenta and the corresponding Kerr-AdS metric was constructed in \cite{Hawking:1998kw}.   The above solution describes the one with equal angular momenta.

An important difference between the black hole and the time machine to be studied in the next subsection is the characteristics of the Killing horizon at $r=r_0$.  The null Killing vector on the horizon, which is a degenerate surface, is given by
\be
\ell = \fft{\partial}{\partial t} + \Omega_+\fft{\partial}{\partial \psi}\,.
\ee
The surface gravity $\kappa$ on the horizon can be obtained from the null Killing vector as
\be
\kappa^2 = - \fft{g^{\mu\nu}\partial_\mu \ell^2\partial_\nu\ell^2}{4\ell^2} = (2\pi T)^2\,.\label{sg}
\ee
The surface gravity defined above with a minus sign implies that the imaginary time is periodic leading to black hole temperature.  It also implies that geodesics do not complete on the event horizon and there is an interior region.

\subsubsection{Time machine}
\label{sec:ads5time}

The thermodynamical quantities (\ref{d5thermo}) imply that for the metric (\ref{d5rot1}) to describe a black hole, we must have that $\mu$ and $\nu$ are both non-negative.  However, the local solution (\ref{d5rot1}) is real as long as we have $\mu\nu\ge 0$.   It is thus of interest to study the global structure of (\ref{d5rot1}) when $\mu$ and $\nu$ are both negative instead.  Let
\be
-\mu=\beta\ge 0\,,\qquad -\nu = \alpha\ge 0\,.\label{minusmunu}
\ee
The solution (\ref{d5rot1}) becomes
\bea
ds_5^2 &=& \fft{dr^2}{f} - \fft{f}{W} dt^2 + \ft14 r^2 W(\sigma_3 + \omega)^2 + \ft14 r^2 d\Omega_2^2\,,\cr
f &=& (1+g^2 r^2) W + \fft{\beta}{r^2},,\qquad
W=1 - \fft{\alpha}{r^4}\,,\qquad \omega = \fft{2\sqrt{\alpha\beta}}{r^4W} dt\,.
\eea
The metric is still asymptotic to AdS$_5$, but with mass and angular momentum given by
\be
M=-\ft18\pi (3\beta+ g^2\alpha)\,,\qquad J=\ft14\pi \sqrt{\alpha\beta}\,.
\ee
Thus the solution has negative mass, with no lower bound.  Naively, one would expect that the metric would then have naked curvature singularity.  This is indeed the case when $\alpha=0$, corresponding to the Schwarzschild-AdS solution with negative mass.  However, if $\alpha$ is non-vanishing,
the manifold described by this metric is smooth, with the local $r=0$ power-law singularity outside the manifold.

As the radial coordinate $r$ decreases from the asymptotic infinity, we come across a special point $r_*=\alpha^{\fft14}$ for which $W=0$.  This is neither coordinate nor curvature singularity, but a velocity of light surface (VLS).  Inside the VLS, we have $g_{\psi\psi}<0$.  In other words, the periodic coordinate $\psi$ becomes time like, giving rise to naked CTCs.  Thus the metric describes a time machine, with the VLS as its boundary.

As $r$ decreases further, at $r=r_0>0$, we have $f(r_0)=0$. This corresponds to a Killing horizon.  The null Killing vector (of zero length) is given by
\bea
&&\ell_1 =\gamma\,\Big(\fft{\partial}{\partial t} +\Omega
\fft{\partial}{\partial\psi}\Big)\,,\nn\\
\gamma = \fft{r_0^2\sqrt{\beta(1 + g^2 r_0^2)}}{\beta + 2 r_0^2(1 + g^2 r_0^2)^2}\,,&&
\Omega = \fft{2\sqrt{\alpha\beta}}{r_0^4W(r_0)} =  \fft{2\sqrt{(1 + g^2r_0^2)(\beta + r_0^2(1+g^2 r_0^2))}}{r_0\sqrt{\beta}}\,.\label{d5ell1}
\eea
It is easy to verify that the surface gravity $\kappa$ defined in (\ref{sg}) is negative, giving rise to imaginary temperature
\be
T=\fft{{\rm i}}{2\pi \gamma}\,.
\ee
It is thus more natural to define a ``Euclidean surface gravity'' $\kappa^{\phantom{X}}_E$ as
\be
\kappa_E^2 = +\fft{g^{\mu\nu}\partial_\mu \ell^2 \partial_\nu \ell^2}{4\ell^2}\,.\label{kappa1}
\ee
The Killing horizon with a real Euclidean surface gravity is called Euclidean pseudo horizon, on which conical singularity can arise potentially.

A simplest example of Euclidean pseudo horizon occurs in two-dimensional flat space written in polar coordinates $ds^2=d\rho^2 + \rho^2 d\phi^2$.  The Killing vector $\ell=\partial_\phi$ is null, i.e. having zero length, in the middle $\rho=0$, with $\kappa^{\phantom{X}}_E=1$. The metric describes Euclidean $\mathbb R^2$ if $\Delta\phi=2\pi$, in which case $\rho=0$ is just an ordinary point in $\mathbb R^2$. If $\Delta \phi\ne 2\pi$, the metric is of a cone with the tip at $\rho=0$.

It is easy to verify that for the Killing vector $\ell_1$, we have $\kappa_E^2=1$. Thus, for the time machine to avoid conic singularity, $\ell_1$ must likewise generate $2\pi$ period.  In other words, it is the real time coordinate rather than the imaginary time coordinate that must be periodic. Once this is imposed, the geodesic completes and spacetime closes off at the Killing horizon.  The local $r=0$ singularity is then outside the manifold.  It should be emphasized that the existence of the Killing horizon $r=r_0$ is independent of whether the cosmological constant $\Lambda=-4g^2$ vanishes or not.  It follows that the above result is applicable also for the asymptotically-flat cases.

     In the standard embedding of AdS$_5$ in the $(4+2)$ flat spacetime, time $t$ in global coordinates is periodic.  The Killing vectors $\ell_0=\fft{1}{g}\fft{\partial}{\partial t}$ and $\ell_2=2\fft{\partial}{\partial \psi}$ both generate $2\pi$ period. It follows from (\ref{d5ell1}) that $(\ell_0,\ell_1,\ell_2)$ are linearly dependent. The consistency requires that coefficients are co-prime integers, namely
\be
n_0 \ell_0 = n_1 \ell_1 + n_2 \ell_2\,.\label{cons1}
\ee
Comparing this to (\ref{d5ell1}), we conclude that the dimensionless parameters $(gr_0,g^2\beta)$ or the original $(g^4\alpha, g^2\beta)$ of the asymptotically-AdS time machines can be expressed in terms of two rational numbers.  Note that the period of $\ell_1$ has to be strictly $2\pi$ to avoid conic singularity.  The period of $\psi$ can be further divided by integer $k$ without introducing singularity, corresponding to AdS$_5/\mathbb Z_k$.  We can also divide or multiply the period $t$ by an integer, corresponding to the quotient or multi-covering of the AdS.

When $g=0$, we have an asymptotically-flat time machine with equal angular momenta.  In this case, the Killing vector $\fft{\partial}{\partial t}$ is not periodic a priori, and hence there is no extra constraint such as (\ref{cons1}).

It is worth commenting that in the case of the rotating black hole discussed in subsection \ref{sec:ad5rot}, the event-horizon topology is 3-sphere.  To be specific, the horizon geometry is a squashed 3-sphere, written as a $U(1)$ bundle over $S^2$.  For the time machine discussed in this section, the Euclidean pseudo horizon is Minkowski signatured, and it is a constant time bundle over $S^2$.  It is also rather counterintuitive that not only the time-machine mass is negative, it has no lower bound.

Finally it is also worth commenting that if the function $f(r)$ had a double zero, there would be no need for periodic identification of the real time coordinate.  The resulting spacetime is called a repulson \cite{Gibbons:1999uv}. None of the examples studied in detail in this paper exhibits repulson-like behavior.

\subsection{$D=2n+1$ time machines with equal angular momenta}

The five-dimensional time machine discussed in the previous subsection can be easily generalized to all $D=2n+1$ dimensions.  We start with the Kerr-AdS black holes with all equal angular momenta.  The Kerr-Schild form was given in \cite{Gibbons:2004uw}.  The Boyer-Lindquist form was presented in \cite{Gibbons:2004ai}, given by
\bea
ds_{2n+1}^2 &=& - \fft{1 + g^2 r^2}{\Xi} dt^2 + \fft{U dr^2}{V-2m} + \fft{r^2 + a^2}{\Xi}
(\sigma^2 + d\Sigma_{n-1}^2) + \fft{2m}{U\Xi^2}(dt +a \sigma)^2\,,\nn\\
\sigma &=& d\psi +A\,,\qquad U= (r^2 + a^2)^{n-1}\,,\qquad
V=\fft{1}{r^2} (1+ g^2 r^2) (r^2 + a^2)^n\,,\label{hdrot}
\eea
where $\Xi=1-a^2 g^2$, and $d\Sigma_{n-1}^2$ is the standard Fubini-Study metric on $\mathbb {CP}^{n-1}$, and the fibre 1-form is $\sigma=d\psi + A$, with $dA=J$ being the K\"ahler form.  The coordinate $\psi$ has period $2\pi$ and the terms $(\sigma^2 + d\Sigma_{n-1}^2)$ in the metric are nothing but the metric on the round sphere $S^{2n-1}$.   The mass and angular momentum are given by
\be
M=\fft{m(2n-\Xi){\cal A}_{2n-1}}{8\pi \Xi^{n+1}}\,,\qquad J= \fft{ma {\cal A}_{2n-1}}{4\pi \Xi^{n+1}}\,,
\ee
where ${\cal A}_{k}$ is the volume of a unit round $S^k$, given by
\be
{\cal A}_k = \fft{2\pi^{\fft12(k+1)}}{\Gamma[\fft12(k+1)]}\,.
\ee
It is instructive to define a new coordinate $\hat r$ that measures the radius of the $S^{2n-1}$ sphere.  Thus we make a coordinate transformation
\be
\fft{r^2+a^2}{\Xi}= \hat r^2\,.
\ee
The metric (\ref{hdrot}) can be written, after dropping the hat, as
\bea
ds_{2n+1}^2 &=& \fft{dr^2}{f} - \fft{f}{W} dt^2 + r^2 W (\sigma + \omega)^2 + r^2 d\Sigma_{n-1}^2\,,\cr
f &=& (1+g^2 r^2) W - \fft{\mu}{r^{2(n-1)}}\,,\qquad W=1 + \fft{\nu}{r^{2n}}\,,\qquad \omega = \fft{\sqrt{\mu\nu}}{r^{2n}+\nu}\, dt\,.
\eea
where the constants $\mu$ and $\nu$ are related to original $(m,a)$ parameters as
\be
a=\sqrt{\fft{\nu}{\mu}}\,,\qquad m=\ft12\mu\Big(1 - \fft{\nu}{\mu}g^2\Big)^{n+1}\,.\label{amn}
\ee
The solutions describe rotating black holes in $D=2n+1$ dimensions when both $(\mu,\nu)$ are positive.  When $n=1$, the metric reduces to the BTZ black hole \cite{Banados:1992wn} after making a trivial coordinate transformation $r^2 + \nu^2 \rightarrow r^2$, and hence all our statements apply also to three dimensions.  When $n=2$, the solution reduces to (\ref{d5rot1}).

As in the previous $D=5$ example, when $(\mu,\nu)$ both take negative values, as in (\ref{minusmunu}), the corresponding metric becomes
\bea
ds_{2n+1}^2 &=& \fft{dr^2}{f} - \fft{f}{W} dt^2 + r^2 W (\sigma + \omega)^2 + r^2 d\Sigma_{n-1}^2\,,\cr
f &=& (1+g^2 r^2) W + \fft{\beta}{r^{2(n-1)}}\,,\qquad W=1 - \fft{\alpha}{r^{2n}}\,,\qquad \omega = \fft{\sqrt{\alpha\beta}}{r^{2n}W}\, dt\,.\label{gendtime}
\eea
The mass and angular momentum are given by
\be
M=-\fft{{\cal A}_{2n-1}}{16\pi} ((2n-1)\beta + g^2 \alpha)\,,\qquad J=\fft{{\cal A}_{2n-1}}{8\pi} \sqrt{\alpha\beta}\,.\label{mj}
\ee
Since $\alpha$ and $\beta$ are positive, the solutions all have negative mass, with no lower bound.

When $\alpha=0$, the solution becomes the Schwarzschild-AdS metric with negative mass, and hence the power-law curvature singularity at $r=0$ is naked.  If on the other hand $\alpha>0$, no matter how small or big, there is a Killing horizon at $r=r_0>0$ where $f(r_0)=0$.  The corresponding null Killing vector takes the form
\bea
\ell = \fft{r_0^n (1 + g^2 r_0^2)}{n r_0^{2n} (1 + g^2 r_0^2)^2 + \beta r_0^2}
\Big(\fft{r_0^2\sqrt{\beta}}{\sqrt{1 + g^2 r_0^2}}\, \fft{\partial}{\partial t} +
\sqrt{r_0^{2n}(1 + g^2 r_0^2) + \beta r_0^2}\, \fft{\partial}{\partial \psi}\Big)\,.
\eea
The overall scaling of the Killing vector is chosen such that the Euclidean surface gravity is unit, as in (\ref{kappa1}).  Consequently, $r=r_0$ is a pseudo horizon where geodesic completes provided that $\ell$ generates $2\pi$ period.  It is easy to see that on the Killing horizon, $g_{\psi\psi}=r_0 W(r_0)<0$.  In fact, naked CTCs arise inside the VLS located $r_*=\alpha^{\fft1{2n}}>r_0$.  The metrics describe smooth time machines with negative mass, provided that $\alpha>0$.  The geometry of the Euclidean pseudo horizon is a constant time bundle over $\mathbb {CP}^n$. The conclusion is valid for both asymptotically-flat ($g^2=0$) or AdS solutions.

\subsection{Time machines with unequal angular momenta}

In $D=2n+1$ dimensions, there can be $n$ independent rotations. We again start with the
Kerr-AdS metrics, but with now arbitrary non-zero rotations.  The metrics were constructed in \cite{Gibbons:2004uw,Gibbons:2004js}.  In analogous notations, they are given by
\bea
ds_{2n+1}^2 &=& - W (1 + g^2 r^2) dt^2 + \fft{U\,dr^2}{V-2m} + \fft{2m}{U} \Big(dt - \sum_{i=1}^n
\fft{a_i \mu_i^2 d\phi_i}{\Xi_i}\Big)^2\nn\\
&&+ \sum_{i=1}^n \fft{r^2 + a_i^2}{\Xi_i} \Big(d\mu_i^2 + \mu_i^2 (d\phi_i + a_i g^2 dt)^2\Big)\nn\\
&&-\fft{g^2}{(1+g^2 r^2)W}\Big(\sum_{i=1}^n \fft{r^2+a_i^2}{\Xi_i} \mu_i d\mu_i\Big)^2\,,
\label{kerradsodd}
\eea
where $\sum_i \mu_i^2=1$ and
\bea
&&\Xi_i=1-a_i^2 g^2\,,\qquad W=\sum_{i=1}^n \fft{\mu_i^2}{\Xi_i}\,,\qquad
U=\sum_{i=1}^n \fft{\mu_i^2}{r^2 + a_i^2}\, \prod_{j=1}^n (r^2 + a_j^2)\,,\nn\\
&& V=\fft{1}{r^2} (1+g^2 r^2) \prod_{i=1}^n (r^2 + a_i^2) = \fft{U}{F}\,,\qquad
F=\fft{r^2}{1+g^2 r^2} \sum_{i=1}^n \fft{\mu_i^2}{r^2 + a_i^2}\,.\label{evenvu}
\eea
For positive $m$ and $\Xi_i$'s, the metrics describe general rotating black holes with mass and angular momenta \cite{Gibbons:2004ai}
\be
D=2n+1:\qquad M = \fft{m\, {\cal A}_{D-2}}{4\pi (\prod_j \Xi_j)} \Big(\sum_{i=1}^n \fft{1}{\Xi_i}-\fft12\Big)\,,\qquad J_i = \fft{m a_i\,{\cal A}_{D-2}}{4\pi \Xi_i (\prod_j \Xi_j)} \,.\label{massodd}
\ee
The event horizon is located at $V-2m=0$.  Indeed the determinant of the sub-manifold of constant $r$ slice has a factor of $(V-2m)$, but Riemann tensor invariants are regular at $V-2m=0$.  These show that $V-2m=0$ gives a degenerate surface, with only coordinate singularity.

We now consider the case with $m<0$.  Naively, one might expect that the solutions have a naked power-law curvature singularity, since it is clear that $V -2m=0$ cannot be satisfied for any real $r$.  However, the fact is that as long as rotating parameters $a_i$'s are all non-vanishing, the geodesic does complete at some Euclidean Killing horizon before reaching the singularity.  To see this, it is important to note that $r=0$ is not a curvature singularity when all $a_i\ne 0$.  Instead curvature singularities are located at $r^2 + a_i^2=0$, together with appropriate $\mu_j$'s for each $i$.  In other words, there is nothing special at $r=0$ and the geodesic can extend further into the $r^2<0$ region. Then it is easy to see that when all $a_i\ne 0$ and $m$ is negative, no matter how small or big $|m|$ is, there exists a pure imaginary $r_0$ with
\be
-a_i^2<r_0^2<0\,,\qquad \hbox{for all $i=1,2,\ldots n,$}
\ee
such that $V-2m=0$.  The $r=r_0$ surface gives rise to a Killing horizon.  It is also straightforward to verify that on the Killing horizon there are CTCs.  For example,
\be
g_{\phi_i\phi_i}\Big|_{\mu_i=1} = \fft{(r_0^2 + a^2)^2}{\Xi_i^2 r_0^2} <0\,,\qquad
\hbox{for all $i=1,2,\ldots n.$}
\ee
This implies that the Killing horizon is a pseudo horizon where geodesic completes provided that the appropriate null Killing vector generates $2\pi$ period, as was discussed in the case of equal angular momenta.  It is also important to note that from the definition of $V$ in (\ref{evenvu}) we conclude that the existence of the Euclidean Killing horizon is independent of whether the cosmological constant parameter $g^2$ vanishes or not.  Hence the conclusion is applicable for both asymptotically-flat or AdS solutions.

It is perhaps convenient to introduce $n+1$ new parameters, $(\mu, \nu_1,\ldots, \nu_n)$, and express $m$ and $a_i$ in terms of these parameters
\be
a_i = \sqrt{\fft{\nu_i}{\mu}}\,,\qquad m=\mu \Big(\prod_{i=1}^n \Xi_i\Big)^{\fft{n+1}{n}}\,,\qquad \Xi_i= 1 - \fft{\nu_i}{\mu} g^2\,.
\ee
The mass and angular momenta become
\be
M=\mu \Big(\prod_j \Xi_j\Big)^{\fft1{n}}\Big(\sum_{i=1}^n \fft1{\Xi_i} - \fft12\Big)\,,\qquad
J_i =\fft{\sqrt{\mu \nu_i}}{\Xi_i} \Big(\prod_j \Xi_j\Big)^{\fft1{n}}\,.\label{mji}
\ee
For the metric to describe a rotating black hole, the parameters $(\mu, \nu_i)$ must be non-negative.  However, the reality condition of the metric only requires that $\mu \nu_i\ge 0$ for all $i$.  Thus we can take all the parameters $(\mu, \nu_i)$ to be negative.  The solutions then describe a general class of time machines with negative mass.  When $\nu_i=\nu$ for all $i$, they reduce to the cohomogeneity-one metrics discussed earlier.

The situation is very different in $D=2n$ even dimensions, for which there are only $(n-1)$ independent orthogonal rotations. The Kerr-AdS metrics are \cite{Gibbons:2004uw,Gibbons:2004js}
\bea
ds_{2n}^2 &=& - W (1 + g^2 r^2) dt^2 + \fft{U\,dr^2}{V-2m} + \fft{2m}{U} \Big(dt - \sum_{i=1}^{n-1}
\fft{a_i \mu_i^2 d\phi_i}{\Xi_i}\Big)^2\nn\\
&&+ \sum_{i=1}^n \fft{r^2 + a_i^2}{\Xi_i} d\mu_i^2 + \sum_{i=1}^{n-1} \fft{r^2 + a_i^2}{\Xi_i} \mu_i^2 (d\phi_i + a_i g^2 dt)^2\nn\\
&&-\fft{g^2}{(1+g^2 r^2)W}\Big(\sum_{i=1}^n \fft{r^2+a_i^2}{\Xi_i} \mu_i d\mu_i\Big)^2\,,
\eea
where $\Xi_i$, $W$ and $U$ take the same for as those in $D=2n+1$ dimensions, except that $a_n=0$ since in $D=2n$ dimensions, there is no azimuthal coordinate $\phi_n$ and hence there is no associated rotation parameter $a_n$.  For positive $m$ and $0<\Xi_i\le 1$, the metrics describe rotating AdS black holes with mass and angular momenta \cite{Gibbons:2004ai}
\be
D=2n:\qquad  M=\fft{m\,{\cal A}_{D-2}}{4\pi (\prod_j \Xi_j)} \sum_{i=1}^{n-1} \fft{1}{\Xi_i}\,,\qquad \qquad J_i = \fft{m a_i\,{\cal A}_{D-2}}{4\pi \Xi_i (\prod_j \Xi_j)}\,.\label{masseven}
\ee
As in the case of odd dimensions, the determinant of the submanifold of constant $r$ slice also has a factor of $(V-2m)$.  However, there is a crucial difference in even dimensions. The function $V$ is now given by
\be
V=\fft{1}{r} (1+g^2 r^2) \prod_{i=1}^{n-1} (r^2 + a_i^2)
\ee
Thus in even dimensions, the coordinate $r$ cannot be purely imaginary.  The $r=0$ is a spacetime power-law curvature singularity.  It follows that for $m<0$, the quantity $(V-2m)$ cannot vanish for any $r>0$ and hence there is no degenerate surface.  The singularity at $r=0$ is thus naked.

\subsection{Further time machines}

For the time machine metric (\ref{gendtime}) to be Einstein, the $\mathbb{CP}^{n-1}$ metric $d\Sigma_{n-1}^2$ can be replaced by any Einstein-K\"alher metrics, at the expense that the asymptotic regions are no longer AdS.  When the base is a direct product of multiple Einstein-K\"ahler spaces, there is a subtlety that the period associated with the fibre 1-form $\sigma$ must be consistent with all these factors of the base \cite{Cvetic:2001ma}. Here we present an example in seven dimensions where $d\Sigma_2^2$ is replaced by the metric of $S^2\times S^2$:
\bea
ds^2 &=& \fft{dr^2}{f} - \fft{f}{W} dt^2 + \ft19 r^2 W\, (\sigma + \omega)^2
+ \ft16 r^2 (d\theta_1^2 +
\sin^2\theta_1 d\phi_1^2 + d\theta_2^2 + \sin^2\theta_2 d\phi_2^2)\,,\cr
\sigma &=& d\psi + \cos\theta_1\, d\phi_1 + \cos\theta_2 d\phi_2\,,\qquad \omega=\fft{\sqrt{\alpha\beta}}{r^6W}\,dt\,.
\eea
The metric is Einstein with $R_{\mu\nu}=-6 g^2 g_{\mu\nu}$, provided that functions $W$ and $f$ are
\be
W = 1 - \fft{\alpha}{r^6}\,,\qquad f= (1 + g^2 r^2) W + \fft{\beta}{9 r^4}\,.
\ee
For this solution, the level surfaces are not of $S^5$ but the $T^{1,1}$ space. The asymptotic region is no longer AdS$_7$, and boundary is $\mathbb T\times T^{1,1}$, instead of $\mathbb T\times S^5$. The Killing horizon and the period of associated null Killing vector can be easily determined.

\section{AdS Solitons with negative mass}
\label{sec:soliton}

In the previous sections, we find that in odd dimensions, when mass is negative, Kerr or Kerr-AdS metrics with all angular momenta turned on describe smooth time machines.  We now consider the possibility of turning off all the angular momenta. There are two ways of doing this. The trivial way leads simply to the Schwarzschild metrics with negative mass.  An alternative limit can lead to static solitons.  Negative mass solitons emerge only when there is a cosmological constant.  When the cosmological constant is zero, the mass vanishes, and we shall study this in section \ref{d=2n}.

\subsection{Cohomogeneity-one metrics}

In the typical way of writing Kerr-AdS black holes, the mass $M$ and angular momentum $J$ are expressed in terms of $m$ and $a$. Turning off the angular momentum parameter $a$ has the effect of reducing the metric to the Schwarzschild black hole. In our parametrization (\ref{d5thermo}), we can have two manifest ways of turning off the angular momentum.  The first is to set $\nu=0$, corresponding to setting $a=0$, giving rise to the usual Schwarzschild black hole.  The alternative is to set $\mu=0$, corresponding to setting $a\rightarrow \infty$, and we have a new non-trivial static configuration.  It follows from (\ref{d5rot1}) that when $\mu=0$ and $\nu=-\alpha$ is negative, we obtain a static soliton in five dimensions.  For general dimensions, we start with the time-machine solution (\ref{gendtime}) and set $\beta=0$, we have
\be
ds^2 = \fft{dr^2}{(1+g^2 r^2)W} - (1+g^2 r^2) dt^2 + r^2 W \sigma^2 + r^2 d\Sigma_{n-1}^2\,,\qquad  W=1 - \fft{\alpha}{r^{2n}}\,,\label{gendsoliton}
\ee
where the 1-form $\sigma$ and the metric $d\Sigma_{n-1}^2$ are defined under (\ref{hdrot}).  For positive $\alpha$, the metric becomes singular at $r=r_0=\alpha^{\fft1{2n}}$.  The absence of the conical singularity requires a specific period for coordinate $\psi$ associated with $\sigma$, namely
\be
\Delta\psi = \fft{2\pi}{n\sqrt{g^2 r_0^2 +1}}\,.
\ee
On the other hand, for the metric $(\sigma^2 + d\Sigma_{n-1}^2)$ to describe a round $S^{2n-1}$, the period for $\psi$ is $2\pi$. If we consider instead more general $S^{2n-1}/\mathbb Z_k$, then we have
\be
\Delta\psi = \fft{2\pi}{k}\,.
\ee
This implies that
\be
g^2r_0^2 = \fft{k^2}{n^2}-1\,,\qquad\Rightarrow\qquad \alpha=\fft{1}{g^{2n}} \Big(\fft{k^2}{n^2}-1\Big)^n\,.
\ee
Thus we have $k>n$ and the mass of the soliton is discretized and negative, given by
\be
M=-\fft{{\cal A}_{2n-1}}{16\pi g^{2(n-1)}\, k} \Big(\fft{k^2}{n^2}-1\Big)^2\,.
\ee
Note that when $n=k$, the solution becomes simply the AdS vacuum and $\Delta\psi=2\pi$.  As $k\rightarrow \infty$, the mass reaches a negative lower bound.

In five dimensions, the metric can be written as
\be
ds^2=-  (g^2r^2+1) dt^2 + \fft{dr^2}{(g^2 r^2 + 1)W} +  \ft14 W r^2\sigma_3^2 + \ft14 r^2 d\Omega_2^2\,,\qquad W=1-\fft{\alpha}{r^4}\,.\label{d5soliton1}
\ee
This solution was first obtained in \cite{Clarkson:2005qx,Clarkson:2006zk}. (The local metric with a positive cosmological constant in Euclidean signature was constructed earlier in \cite{Lu:2004ya}, which can describe smooth compact manifolds.) When the cosmological constant vanish, i.e. $g=0$, the metric is a direct product of time and the EH instanton \cite{Eguchi:1978xp}.  The global analysis for (\ref{d5soliton1}) was performed and descretized negative mass was obtained.  The negativeness of the soliton mass was demonstrated also using holographic stress energy in  \cite{Clarkson:2005qx,Clarkson:2006zk} and the Noether procedure \cite{Cebeci:2006hx}.  In our approach, the solutions were obtained in some special limit of Kerr-AdS metrics, and hence the mass formula is a direct consequence of that of Kerr-AdS black holes.

\subsection{First-order equations without superpotential}

It is well-known that EH instanton can be obtained from a set of first-order equations associated with some superpotential.  It turns out that the solitons (\ref{gendsoliton}) in general odd dimensions can also arise from a set of first-order equations.  For simplicity, we demonstrate this explicitly in five dimensions and show that the static soliton (\ref{d5soliton1}) can arise as solutions of some first-order differential equations, instead of Einstein's second-order equations of motion.  However, we also demonstrate that there is no superpotential associated with this first-order system.

The most general ansatz for static metrics with the $SU(2)\times U(1)$ isometry of squashed $S^3$ is
\be
ds^2 = d\rho^2 - a^2 dt^2 + b^2 \sigma_3^2 + c^2 d\Omega_2^2\,,
\ee
where the metric $d\Omega_2$ and 1-form $\sigma_3$ are given in (\ref{s2sig3}) and $(a,b,c)$ are functions of the radial coordinate $\rho$. A dot denotes a derivative with respect to $\rho$.  For the metric to be Einstein with $R_{\mu\nu}+4g^2 g_{\mu\nu} =0$, the $(a,b,c)$ functions satisfy
\bea
-\fft{\ddot a}{a}  - \fft{\ddot b}{b} - \fft{2\ddot c}{c} &=& 4g^2\,,\nn\\
\fft{\ddot a}{a} + \fft{2 \dot a \dot c}{a c} + \fft{\dot a \dot b}{ab} &=& 4g^2\,,\nn\\
\fft{\ddot b}{b} + \fft{2 \dot b \dot c}{b c} + \fft{\dot a \dot b}{ab} - \fft{b^2}{2c^4} &=&
4g^2\,,\nn\\
\fft{\ddot c}{c} + \fft{\dot c^2}{c^2} + \fft{\dot b\dot c}{b c} + \fft{\dot a\dot c}{a c} -
\fft{1}{c^2} + \fft{b^2}{2 c^4} &=& 4g^2\,,
\eea
We find that there exists a set of first-order equations that can solve the above second-order equations of motion, namely
\be
\dot a = \fft{2 g^2 a b}{\sqrt{1 + 4g^2 c^2}}\,,\qquad
\dot b = \fft{(2c^2-b^2)\sqrt{1 + 4g^2 c^2}}{2c^2}\,,\qquad
\dot c =\fft{b\sqrt{1 + 4g^2 c^2}}{2c}\,.\label{fo}
\ee
It is easy to verify that these first-order equations yield precisely the soliton solution (\ref{d5soliton1}).

We now demonstrate that this first-order system is not associated with any superpotential.
To see this, it is convenient to define a new radial coordinate $\eta$, related to $\rho$ by $d\eta=a b c^2 d\rho$. In this system, the effective Lagrangian is given by $L=T - V$ where
\be
T=\fft{2a'b'}{2ab} + \fft{4a'c'}{ac} + \fft{4b'c'}{bc} + \fft{2c'^2}{c^2}\,,\qquad
V=\ft12 a^2 b^2 (b^2 - 4 c^2 - 24 g^2 c^4)\,.
\ee
Here a prime denotes a derivative with respect to $\eta$.  Thus we have $ab c^2 \dot f = f'$ for any function $f$. Following the prescription of \cite{Cvetic:2000db},
we may define $X^i = (a,b,c)$ and write the kinetic energy as $T=\ft12 g_{ij} X'^i X'^j$.  If there would exist a superpotential $U=U(a,b,c)$ such that
\be
V=\ft12 g^{ij}\,\fft{\partial U}{\partial X^i} \fft{\partial U}{\partial X^j}\,,
\ee
then there would be a first-order system
\be
ab c^2\,\dot X^i=g^{ij} \fft{\partial U}{\partial X^j}\,.
\ee
Substituting the first-order equations (\ref{fo}) into the above, and we find
\bea
\fft{\partial U}{\partial a} &=& (b^2 + 2 c^2) \sqrt{1 + 4c^2 g^2}\,,\nn\\
\fft{\partial U}{\partial b} &=& \fft{2 a b (1 + 6 c^2 g^2)}{\sqrt{1 + 4 c^2 g^2}}\,,\nn\\
\fft{\partial U}{\partial c} &=& \fft{4ac\big(1 + 2(b^2 + 2 c^2) g^2\big)}{\sqrt{1+4c^2g^2}}\,.
\eea
It is easy to verify that the above equations do not satisfy the integrability condition unless
$g=0$, in which case we have $U=a(b^2+2c^2)$.  This is precisely the superpotential for generating the EH instanton.  For non-vanishing $g$, on the other hand, although we have the first-order equation (\ref{fo}), there is no superpotential associated with the system.

\subsection{Higher-cohomogeneity solitons}

\subsubsection{$D=5$}
\label{d5cohomo2}

The local solutions of the static solitons were obtained by taking a limit from Kerr-AdS metrics such that the two equal angular momenta vanish whilst the mass is non-vanishing.  Such a limit typically leads to the Schwarzschild-AdS black holes.  However, as we have shown in the previous subsection, there can be an alternative limit.  This new limit can be performed also for the Kerr-AdS metrics with two general angular momenta.  We start with the five-dimensional Kerr-AdS black hole constructed in \cite{Hawking:1998kw}, which involves three parameters, $(m,a,b)$.  Since we shall use the exact convention for the metric presented in \cite{Hawking:1998kw}, we shall not give it here. The mass and angular momenta are given by
\cite{Gibbons:2004ai}
\bea
M=\fft{\pi m (2\Xi_a + 2\Xi_b - \Xi_a \Xi_b)}{4\Xi_a^2 \Xi_b^2}\,,\qquad
J_a = \fft{\pi m a}{2\Xi_a^2 \Xi_b}\,,\qquad J_b=\fft{\pi m b}{2\Xi_b^2 \Xi_a}\,,
\eea
where $\Xi_a=1-a^2 g^2$ and $\Xi_b = 1- b^2 g^2$.  Setting $a=b=0$ turns off the angular momenta and gives rise to the Schwarzschild-AdS black hole of mass $M=\ft34\pi m$.  We would like instead to send $a,b,m$ to infinity such that we have $J_a,J_b\rightarrow0$ while keeping $M$ finite and non-vanishing.  To be specific, we scale the parameters
\be
a= \lambda \tilde a\,,\qquad b=\lambda \tilde b\,\qquad
m= \ft12 \lambda^4 g^6 \tilde a^2 \tilde b^2 \alpha\,,\label{d5limit}
\ee
and then send $\lambda\rightarrow \infty$.  The mass and angular momenta become
\be
M=-\ft18\pi g^2\, \alpha\,,\qquad J_a=J_b=0\,.
\ee
Thus we arrive at a static solution with negative mass.  Making a coordinate transformation $r={\rm i} \lambda \tilde r$, (with $\lambda\rightarrow \infty$,), the Kerr-AdS metric of \cite{Hawking:1998kw} becomes
\bea
ds_5^2 &=& - \fft{r^2 \Delta_\theta dt^2}{a^2 b^2} + \fft{\rho^2 d\theta^2}{g^2\Delta_\theta} + \fft{\rho^2 dr^2}{\Delta_r} + \fft{\Delta_r}{g^2\rho^2} \Big( \sin^2\theta \fft{d\phi_1}{a g} + \cos^2\theta \fft{d\phi_2}{bg}\Big)^2 \cr
&&+ \fft{\sin^2\theta\cos^2\theta}{\rho^2} \Big(
(r^2-a^2) \fft{d\phi_1}{ag} - (r^2-b^2) \fft{d\phi_2}{bg}\Big)^2\,,\nn\\
\Delta_r &=& g^2 \Big((r^2-a^2)(r^2-b^2) - a^2 b^2 g^4\alpha\Big)\,,\qquad
\Delta_\theta=a^2\cos^2\theta + b^2\sin^2\theta\,,\nn\\
\rho^2 &=& r^2 - \Delta_\theta\,.\label{absoliton}
\eea
Here we have dropped all the tildes.  If we set the parameter $\alpha=0$, the metric is exact AdS. At large $r$, the $\alpha$-term in the metric can be neglected.  Thus the metric with non-vanishing $\alpha$ is asymptotic to the AdS spacetime.  The Riemann tensor squared is given by
\be
\hbox{Riem}^2=40g^4 + \fft{24\alpha^2 a^4b^4 g^{12}}{\rho^{12}}(r^2 + 3a^2\cos^2\theta + 3b^2\sin^2\theta)(3r^2 + a^2 \cos^2\theta + b^2\sin^2\theta)\,,
\ee
indicating the metric has a curvature singularity at $\rho=0$. We shall see presently that this curvature singularity is outside the soliton manifold. When $b=a$, we make a further coordinate transformation
\be
\phi_1=\ft12 (\psi-\phi)\,,\qquad \phi_2 = \ft12 (\psi +\phi)\,,\qquad \fft{r^2-a^2}{a^2 g^2}\rightarrow r^2\,,\qquad \theta\rightarrow \ft12 \theta\,.
\ee
the metric (\ref{absoliton}) reduces precisely to (\ref{d5soliton1}).

The power-law curvature singularity $\rho=0$ can be avoided for $\alpha>0$ because there is a Euclidean Killing horizon at $r=r_0>{\rm max}\{a,b\}$ for which $\Delta_r(r_0)=0$.  The condition for existing such $r_0$ is that
\be
\alpha > - \fft{(a^2-b^2)^2}{4a^2 b^2 g^4}\,,\qquad \rightarrow \qquad
M< \fft{(a^2-b^2)^2}{32 \pi a^2 b^2 g^2}\,.\label{massup}
\ee
If the inequality is saturated, $\Delta(r)$ has a double zero and the metric has a power-law curvature singularity at $r=\sqrt{(a^2 + b^2)/2}$ and $\theta=\pi/4$.  It is of interest to note that not only the mass can be negative, but also there is no lower bound.

The metric (\ref{absoliton}) is degenerated at three places with three null Killing vectors
\bea
\theta=\ft12\pi: &&\quad \ell_1 = \fft{\partial}{\partial \phi_1}\,,\cr
\theta=0:&& \quad \ell_2 = \fft{\partial}{\partial \phi_2}\,,\cr
r=r_0: &&\quad \ell_3 = \fft{1}{r_0 (2r_0^2 - a^2-b^2)} \Big(a (r_0^2-b^2) \fft{\partial}{\partial \phi_1} + b (r_0^2 - a^2)\fft{\partial}{\partial \phi_2}\Big)\,.
\eea
All three Killing vectors must generate $2\pi$ period in order to avoid conical singularity. On the other hand, $\ell_3, \ell_1, \ell_2$ are linearly dependent. Therefore they must satisfy
\be
n_3 \ell_3 = n_1 \ell_1 + n_2\ell_2\,,\qquad \hbox{where}\qquad
n_1,n_2,n_3\,\, \hbox{are co-prime integers}\label{n1n2n3}
\ee
Thus
\be
r_0\, \sqrt{n_2 x-n_1} = b \sqrt{x(n_2-n_1 x)}\,,\qquad
n_3 = \fft{n_1 + n_2 x}{b x} r_0\,,
\ee
where $x\equiv \fft{a}{b}$. With this parametrization, the mass parameter $m$ becomes
\be
\alpha= \fft{n_1 n_2 (x^2-1)^2}{g^4 x (n_2 x -n_1)^2}\,.
\ee
We shall not classify all possible $(n_1,n_2,n_3)$ that could arise. Instead, we present an example: $(n_1,n_2,n_3)=(1,2,5)$, which implies that $a=0.713 b$ and $m=3.77/g^4$ and $r_0=1.47 b$.

In fact there is a further subtle conic singularity.  As was noted in \cite{Martelli:2005wy}, the Killing vectors $(\ell_1,\ell_3)$ and $(\ell_2,\ell_3)$ can be simultaneously null at $(r,\theta)=(r_0,\ft12\pi)$ or $(r_0,0)$ respectively.  In Euclidean signature, any linear combination of two null Killing vectors is also null, and hence $ (n_3 \ell_3 - n_1\ell_1) $ or $(n_3 \ell_3 - n_2 \ell_2)$ must generate also $2\pi$ period.  The consistency then requires that $n_1=n_2=1$.  This corresponds to the cohomogeneity-one solutions with $a=b$, discussed earlier.  The example of $(n_1,n_2,n_3)=(1,2,5)$ still have a conic singularity of ADE type at $(r,\theta)=(r_0,\theta=\ft12\pi)$. The cone is not 2-dimensional like $d\rho^2 + \rho^2 d\phi^2$, but four dimensional with $d\rho^2 + \rho^2 d\tilde \Omega^2$, where $d\tilde\Omega^2$ is not a round $S^3$, but a lens space.  For the specific $(n_1,n_2,n_3)=(1,2,5)$ example, the lens space is $S^3/\mathbb Z_2$, giving rise to the $\mathbb R^4/\mathbb Z_2$ orbifold singularity.  Such singularity can be resolved by an EH instanton whose asymptotic region is precisely $\mathbb R^4/\mathbb Z_2$ \cite{Page:1979zu}.

\subsubsection{$D=2n+1$}

We obtain some non-trivial static soliton solutions from Kerr-AdS$_5$ metrics by taking some appropriate limit (\ref{d5limit}).  Under this limit, all angular momenta vanish, whilst the mass becomes a finite negative number.  The resulting metric is specified by three integration constants.  This procedure can be generalized to general odd dimensions.  Kerr-AdS metrics in general dimensions were constructed in \cite{Gibbons:2004uw,Gibbons:2004js}, involving a mass parameter $m$ and $n=[(D-1)/2]$ parameters $a_i$ for angular momenta.  The mass and angular momenta are given in (\ref{massodd}) and (\ref{masseven}) for odd and even dimensions.

We can turn off the angular momenta by setting $a_i=0$, leading to the Schwarzschild-AdS black hole. We now would like to turn off the angular momenta while keeping mass constant by sending $a_i\rightarrow \infty$ and hence $\Xi_i\rightarrow -\infty$.  This is not possible in even dimensions because of the relation
\be
\sum_{i=1}^n \fft{J_i}{a_i}=M\,,
\ee
which can be derived from (\ref{masseven}). In odd dimensions, this can be achieved indeed, because there is the less convergent ``$\ft12$'' term in (\ref{massodd}) in this limit.  Thus, following the $D=5$ example, we make the constant scaling of the parameters
\be
a_i=\lambda \tilde a_i\,,\qquad m = \ft12 (-\lambda^2)^n g^2\alpha \prod_i^n (\tilde a_i g)^2\,,
\ee
and then take the $\lambda\rightarrow \infty$ limit.  Dropping the tildes, we find that the Kerr-AdS metric (\ref{kerradsodd}) becomes
\bea
ds_{2n+1}^2 &=&-r^2 \Big(\sum_{i=1}^n \fft{\mu_i^2}{a_i^2}\Big)\,dt^2 + \fft{X}{Y} dr^2 +
\sum_{i=1}^n \fft{r^2-a_i^2}{a_i^2g^2} (d\mu^2_i + \mu_i^2 d\phi_i^2)\nn\\
&& - \fft{1}{r^2Z} \Big(\sum_{i=1}^n \fft{r^2-a_i^2}{a_i^2 g^2} \mu_i d\mu_i\Big)^2
-\fft{\alpha g^2}{X} (\prod_{i=1}^n (a_i g)^2)\, \Big(\sum_{i=1}^n \fft{\mu_i^2 d\phi_i}{a_i g^2} \Big)^2\,,\label{gendsoliton2}
\eea
where $\sum_i^n u_i^2=1$ and
\bea
X &=& \Big(\prod_{i=1}^n (r^2-a_i^2)\Big) \sum_{i=1}^n \fft{\mu_i^2}{r^2-a_i^2}\,,\nn\\
Y &=& g^2 \prod_{i=1}^n (r^2 -a_i^2) - g^2\alpha \prod_{i=1}^n (a_i g)^2\,,\nn\\
Z &=& \sum_{i=1}^n \fft{\mu_i^2}{a_i^2 g^2}\,.
\eea
The metrics are static and hence there is no angular momentum.  The mass of the soliton is negative,
given by
\be
M=-\fft{{\cal A}_{D-2}}{16\pi} g^2\alpha\,.
\ee
We shall not discuss the global structure of this general class of AdS solitons in this paper.

\section{Ricci-flat instantons in $D=2n$ dimensions}
\label{d=2n}

In the previous section, we obtained large classes of static AdS solitons in $D=2n+1$ dimensions. For the cohomogeneity-one metrics (\ref{gendsoliton}), it can be easily seen that in the $g=0$ limit, the resulting spacetime is a direct product of time and the $D=2n$ gravitational instanton that is a higher-dimensional generalization of the EH instanton.  The metric (\ref{gendsoliton}) was generalized to multi-cohomogeneity metrics (\ref{absoliton}) in $D=5$ and (\ref{gendsoliton2}) in $D=2n+1$.  In this section, we perform a further $g=0$ limit on (\ref{absoliton}) and (\ref{gendsoliton2}) and obtain Ricci-flat gravitational instantons in $D=2n$ dimensions.

\subsection{$D=4$}

We start with the five-dimensional Einstein metric (\ref{absoliton}) and reparameterize the $(a,b,\alpha)$ constants as
\be
a^2 =a_0^2(1 - g^2 \beta^2)\,,\qquad b^2 = a_0^2 (1+ g^2\beta^2)\,,\qquad
\alpha\rightarrow \alpha -\beta^4\,.
\ee
Making first the coordinate transformation,
\be
\phi_1=\ft12 (\psi-\phi)\,,\qquad \phi_2 = \ft12 (\psi +\phi)\,,\qquad \fft{r^2-a_0^2}{a_0^2 g^2}\rightarrow r^2\,,\qquad \theta\rightarrow \ft12 \theta\,.
\ee
and then sending $g\rightarrow 0$, we obtain a smooth limit of (\ref{absoliton}), whose $D=4$ spatial section is
\bea
ds_4^2 &=& \fft{U dr^2}{W} + \fft{W}{4U} r^2 (d\psi +\cos\theta d\phi)^2+ \ft14 r^2 \Big(U d\theta^2 + \fft{1}{U} \sin^2\theta\, \big(d\phi - \fft{\beta^2}{r^2} d\psi\big)^2\Big)\,,\nn\\
W &=& 1 - \fft{\alpha}{r^4}\,,\qquad U=1 + \fft{\beta^2\cos\theta}{r^2}\,.
\eea
Note that the constant $a_0$ is trivial and drops out. The metric is Ricci-flat and K\"ahler.  The K\"ahler structure can be easily seen by constructing the covariant K\"ahler 2-form
\be
J=e^0\wedge e^3+ e^1\wedge e^2\,,
\ee
where the vielbein are
\bea
e^0 &=& \sqrt{\fft{U}{W}}\, dr\,,\qquad
e^3 = \sqrt{\fft{W}{4U}} r\, (d\psi + \cos\theta d\phi)\,,\nn\\
e^1 &=& \ft12 r \sqrt{U} d\theta\,,\qquad
e^2 =-\fft{r}{2\sqrt{U}} \sin\theta\, \big(d\phi - \fft{\beta^2}{r^2}d\psi\big)\,.
\eea
Thus the metric is the Ricci-flat and BPS limit of the general Plebanski solutions \cite{pleb}. When $\beta=0$, the metric is the EH instanton.  For $\beta\ne0$, the curvature singularity is located at $U=0$, which can be avoided if $\beta< \alpha^{\fft14}$.  There are three degenerate surfaces whose null Killing vectors are
\bea
\theta=0:&&\qquad \ell_1= \fft{\partial}{\partial \psi} - \fft{\partial}{\partial\phi}\,,\nn\\
\theta=\pi:&&\qquad \ell_2=\fft{\partial}{\partial \psi} + \fft{\partial}{\partial\phi}\,,\nn\\
r=\alpha^{\fft14}:&&\qquad \ell_3=\fft{\partial}{\partial\psi} + \fft{\beta^2}{\sqrt{\alpha}}
\fft{\partial}{\partial\phi}\,,
\eea
all of which have unit Euclidean surface gravity $\kappa_{E}^{\phantom{X}}$.  When $\beta^2/\sqrt{\alpha}=p/q<1$ is a rational number, then we have
\be
2q \ell_3 = (q-p) \ell_1 + (q+p)\ell_2\,.\label{pqcons}
\ee
It follows from  (\ref{n1n2n3}) that $n_1=(q-p)$, $n_2=q+p$ and $n_3=2q$.  Further regularity conditions follow the same procedure described in subsection \ref{d5cohomo2}.  The existence of the ADE-type conical codimension-3 singularity, albeit may be resolved, suggests that these metrics are outside the classes of Gibbons-Hawking instantons \cite{Gibbons:1979zt,Gibbons:1979xm}.  Furthermore, the relation (\ref{pqcons}) implies that the asymptotic regions are cones of more general lens spaces, rather than the $S^3/\mathbb Z_{k+1}$ for $k$ number of EH instantons.

\subsection{$D=2n$}

For general even dimensions, we start with the Einstein metric (\ref{gendsoliton2}) and reparameterize the integration constants
\be
a_i^2 = a_0^2 (1 + g^2 b_i^2)\,,\qquad \sum_{i=1}^n b_i^2=0\,.
\ee
(Note that the resulting metric is real as long as $b_i$'s are either real or purely imaginary numbers.)  Making a coordinate transformation
\be
\fft{r^2-a_0^2}{a_0^2 g^2} \rightarrow r^2\,,
\ee
and then sending the cosmological constant parameter $g$ to zero, we find that the metric (\ref{gendsoliton2}) has a smooth limit and it is a direct product of time and a $D=2n$ Ricci-flat metric
\bea
ds_{2n}^2 &=& \fft{U}{W} r^2 dr^2 + \sum_{i=1}^n (r^2 - b_i^2) (d\mu_i^2 + \mu_i^2 d\phi_i^2) - \fft{\alpha}{U} (\sum_{i=1}^n \mu_i^2 d\phi_i)^2\,,\nn\\
W &=& \prod_{i=1}^n (r^2 - b_i^2) - \alpha\,,\qquad U=\prod_{i=1}^n (r^2-b_i^2) \sum_{j=1}^n \fft{\mu_j^2}{r^2-b_j^2}\,.
\eea
The curvature power-law singularity is at $U=0$, which can be avoided if the geodesics complete in the $r$ region $r\in [r_0,\infty)$ where $W(r_0)=0$.  There are $n+1$ degenerate surfaces and the corresponding null Killing vectors are
\bea
r=r_0: &&\qquad  \ell_0 =\sum_{i=1}^n \fft{\prod_j (r_0^2-b_j^2)}{P(r_0)(r_0^2-b_i^2)} \,\fft{\partial}{\partial\phi_i}\,,\nn\\
\mu_i=0: &&\qquad \ell_i =\fft{\partial}{\partial\phi_i}\,,\qquad i=1,2,\ldots,n\,.
\eea
Here $P(r_0)$ is an $2(n-1)$-order polynomial of $r_0$ with the leading term as $nr_0^{2(n-1)}$.  For example, we have $P=2r_0^2$ for $n=2$ and $P=3r_0^4 + \ft12 (b_1^4 + b_2^4 + b_3^4)$ for $n=3$. All these Killing vectors are scaled such that they have unit Euclidean surface gravity. Therefore they must all generate $2\pi$ period to avoid conical singularities.  We shall study the global structure of these metrics in a future publication since these massless solutions are outside the scope of this paper. We expect all these metrics are Ricci-flat K\"aher, locally the same as those BPS limits of Kerr-AdS-NUT solutions obtained in \cite{Chen:2006xh}. In particular when all $b_i$'s vanish, the metric reduces to the spatial section of (\ref{gendsoliton}) with $g=0$, which is on a smooth manifold of Ricci-flat K\"ahler.  In general, the metrics are cones of Einstein-Sasaki spaces in the asymptotic regions and isolated examples smooth metrics with higher cohomogeneity were found in \cite{Oota:2006pm,Lu:2006cw}.

\section{Euclidean AdS solitons with negative mass}
\label{sec:instanton}

For a Schwarzschild black hole, we can Wick rotate the time coordinate $t$ so that the solution becomes a Euclidean-signatured soliton that is asymptotic to $\mathbb R^{D-1}\times S$.  For Kerr metrics or Kerr-(A)dS metrics, the reality condition requires that the rotation parameters $a_i$ become pure imaginary after the Wick rotation.  In other words, we must have
\be
t={\rm i} \,\tau\,,\qquad a_i \rightarrow {\rm i}\,a_i\,.
\ee
For positive cosmological constant, the resulting metric becomes compact and the absence of conical singularities on the Euclidean Killing horizons put strong constraints on the parameter spaces.  Consequently the complete manifolds are classified by a set of integer values.  This was done in general for Kerr-dS metrics in \cite{Gibbons:2004uw}.  Einstein-Sasaki metrics $Y^{pq}$ \cite{Gauntlett:2004yd} and more general $L^{pqr}$ \cite{Cvetic:2005ft,Cvetic:2005vk} can also be constructed in this procedure.

In this section, we consider a negative or zero cosmological constant, and hence the manifolds are non-compact.  An interesting phenomenon occurs in odd dimensions.  Before the Wick rotation, we have $0<\Xi_i\le 1$ for $i$, it follows from (\ref{masseven}) and (\ref{massodd}) that the mass are positive definite, provided that $m>0$.  Under $a_i\rightarrow {\rm i} a_i$, we have
\be
\Xi_i=1 + a_i^2 g^2\ge 1\,.
\ee
It follows from (\ref{massodd}) that the mass for even dimensions remain positive definite.  However, in odd dimensions, the mass for Euclidean solitons can become negative provided that none of the $a_i$ vanishes and they are all sufficiently large so that
\be
\sum_{i=1}^{n} \fft{1}{\Xi_i} < \ft12\,.
\ee
When the above bound is saturated, we obtain a massless soliton.  Of course, when the above bound is violated, we get solitons with positive mass.  It is clear that the cosmological constant $g^2$ plays a crucial role in the above discussion and hence the solitons can only have negative mass for asymptotic AdS spacetimes.

To demonstrate this explicitly, we start with the cohomogeneity-one Kerr-AdS metric with all equal angular momenta.  In five dimensions, the metric can be written as (\ref{d5rot1}).  We can perform Wick rotation and choose the parameters
\be
t={\rm i} \tau\,,\qquad \mu=b>0\,,\qquad \nu=-a<0\,.
\ee
In general $D=2n+1$ dimensions, we can start with (\ref{gendtime}) and perform Wick rotation and set $\beta=-b$ and $\alpha=a$, we find that the Euclidean soliton is
\bea
ds^2 &=& \fft{dr^2}{f} + \fft{f}{W} d\tau^2 + r^2 W \big(\sigma + \fft{\sqrt{ab}}{r^{2n} W} d\tau\big)^2 + r^2 d\Sigma_{n-1}^2\,,\nn\\
f &=& (1 + g^2 r^2) W - \fft{b}{r^{2(n-1)}}\,,\qquad W= 1 - \fft{a}{r^{2n}}\,,
\eea
where $a>0$ and $b>0$.  If follows from (\ref{mj}) that we can define the ``Euclidean mass'', given by
\be
M=\fft{{\cal A}_{2n-1}}{16\pi} \Big( (2n-1) b - g^2 a\Big)\,.
\ee
The metric has a Killing horizon at $r=r_0$ which is the largest real root of $f$.  We can express $b$ in terms of $(r_0,a)$, given by
\be
b=\fft{1}{r_0^2} (1 + g^2 r_0^2) (r_0^{2n} - a)\,.
\ee
The coordinate $\tau$ then must have period
\be
\Delta \tau = \fft{4\pi \sqrt{W(r_0)}}{f'(r_0)}\,,
\ee
provided that we let $\sigma\rightarrow \sigma -\sqrt{ab}/(r_0 W(r_0))\,d\tau$.  Note that
the condition $b\ge 0$ implies $a\le r_0^{2(n-1)}$. It follows that there is a lower bound of the mass
\be
M\ge -\fft{{\cal A}_{2n-1}}{16\pi} g^2 r_0^{2n}\,.
\ee
(This should be compared to the Minkowski-signatured AdS soliton, whose mass has an upper bound (\ref{massup}).) Thus mass can be also negative for Euclidean AdS solitons in odd dimensions.  In particular, the parameter region
\be
\fft{(2n-1)(1 + g^2 r_0^2)}{2n(1+ g^2 r_0^2) -1} \le \fft{a}{r_0^{2n}} \le 1
\ee
corresponds to $0\ge M\ge - g^2 r_0^{2n}$. Thus when the lower bound is saturated we have a massless soliton. When $a$ is sufficiently small so that the above lower bound is violated, then the mass becomes positive.  It is worth commenting that in the extremal case where $f$ has a double root, the mass is positive.

The existence of negative mass in Euclidean-signatured space is not uncommon.  The Atiyah-Hitchin metric is a solution of the Euclidean Taub-NUT with negative mass \cite{Atiyah:1985dv,Gibbons:1986df}, where the asymptotic region is $\mathbb R^3\times S$.  Analogous solutions exist also in higher dimensions \cite{Cvetic:2001sr}.

\section{Time machine with a dipole charge}

In the previous sections, we have focused on the Einstein metrics with $R_{\mu\nu}=-2n g^2\, g_{\mu\nu}$ in $D=2n+1$ dimensions.  We now consider charged rotating solutions.
Exact solutions of charged Kerr-AdS black holes in higher dimensions are known only in supergravities.  In five dimensions, notable examples include ones in supergravities \cite{Cvetic:1996xz} and gauged supergravities \cite{Chong:2005hr,Wu:2011gq}.  BPS solutions are somewhat simpler and global analysis indicates that both black holes or time machines can arise, see e.g.~\cite{Breckenridge:1996is,Klemm:2000vn,Klemm:2000gh,Gauntlett:2002nw,
Gutowski:2004ez,Cvetic:2005zi}. In this section we consider the charged Kerr-AdS black hole in minimal gauged supergravity in five dimensions \cite{Chong:2005hr}.  Soliton limits of this solution were studied in \cite{Compere:2009iy}.  We consider a very different limit such that the resulting solution carries no electric charge, but only the magnetic dipole charge.

\subsection{Asymptotic to AdS$_5$}

We follow the same parametrization of \cite{Chong:2005hr}, and make redefinitions on the parameters as well as the coordinate $r$
\bea
a = \lambda \tilde a \,,\qquad b= \lambda \tilde b \,,\qquad m = \ft12 \lambda^4 \tilde a^2 \tilde b^2 g^6 \alpha \,,\qquad q =- \lambda^3 g\tilde q  \,,\qquad r= {\rm i} \lambda \tilde r \,. \label{d5lim}
\eea
We then send the scaling parameter $\lambda\rightarrow \infty$ and find that the charged Kerr-AdS metric of \cite{Chong:2005hr} has a smooth limit.  Dropping all the tildes, the solution can be written as
\bea
ds^2 &=& \frac{\rho^2}{\Delta_r}dr^2+\frac{\rho^2}{g^2 \Delta_\theta} d\theta^2 -\frac{\Delta_\theta (a b q \omega - g r^2 \rho^2 dt)^2}{a^2 b^2 g^2 r^2 \rho^4}+\frac{\Delta_r}{g^2 \rho^2} \omega^2\nn \\
&& + \frac{\sin^2\theta \cos^2\theta}{\rho^2} \Big((r^2-a^2) \frac{d\phi_1}{a g} - (r^2 - b^2) \frac{d\phi_2}{b g} \Big)^2\,,\nn\\
A &=& \fft{\sqrt3\,q}{\rho^2}\, \omega\,,\qquad \omega = \frac{\sin^2\theta}{a g}d\phi_1 + \frac{\cos^2 \theta}{b g} d\phi_2\,,
\eea
where
\bea
\Delta_r &=& g ^2 \Big((r^2 - a^2)(r^2 - b^2) - a^2 b^2 g^4 \alpha-
\fft{q^2}{r^2}\Big)\,, \nn\\
\Delta_\theta &=& a^2 \cos^2\theta+b^2 \sin^2\theta\,,\qquad
\rho^2 = r^2-\Delta_\theta\,.
\eea
Under the limit (\ref{d5lim}) with $\lambda\rightarrow \infty$, the electric charge vanishes, but mass, angular momenta are given by
\be
M = -\fft18 \pi g^2 \alpha \,,\qquad J_a =\fft{\pi q}{4 a b^2 g^3} \,,\qquad
J_b =\fft{\pi q}{4 a^2 b g^3} \,.\label{mjajb}
\ee
The rotating is generated by the magnetic flux whose strength is characterized by the parameter $q$.  When $q=0$, the solution becomes static and reduces to (\ref{absoliton}).  There is a Euclidean Killing horizon at $r=r_0$ for which $\Delta_r(r_0)=0$ and the corresponding null vector
\be
\ell= \fft{ab q r_0}{q^2 + r_0^4(2r_0^2 - a^2 - b^2)}\Big(
\fft{1}{g} \fft{\partial}{\partial t} +\fft{r_0^2 (r_0^2-b^2)}{b q} \fft{\partial}{\partial \phi_1} + \fft{r_0^2(r_0^2-a^2)}{a q} \fft{\partial}{\partial \phi_2}\Big)\,,
\ee
must generate $2\pi$ period to avoid conical singularity.  The existence of naked CTCs can be seen, for example, from $g_{\phi_1\phi_1}$ which is obviously negative at $r=r_0$ and $\theta=\ft12\pi$.  For non-vanshing $q$, the existence of the Killing horizon is independent of the value and sign of the constant $\alpha$. It follows (\ref{mjajb}) that the mass can be either positive or negative, without upper or lower bounds. On the Killing horizon, there is a magnetic dipole charge, given by
\be
{\cal D} = \fft{1}{8} \int_{r_0} F=\ft18\sqrt3 q\Big(\fft{\Delta\phi_1}{ag (r_0^2-b^2)} +
\fft{\Delta \phi_2}{b g(r_0^2 - a^2)}\Big)\,.
\ee
The reason why dipole charge is consistent with a time machine is that the topology of the Killing horizon is a time bundle over $S^2$.

The solution becomes much simpler when $b=a$. Making coordinate transformations
\bea
\phi_1 = \fft12 (\psi - \phi) \,, \quad \phi_2 = \fft12 (\psi+ \phi) \,, \quad \frac{r^2 - a^2}{a^2 g^2} = \tilde{r}^2 \,, \quad \theta = \fft12 \tilde{\theta} \,,\qquad
q=a^3 g^3 \tilde q\,,
\eea
and then dropping the tildes, we have
\bea
ds^2 &=& - g^2 r^2 dt^2 - \Big(dt + \fft{q}{2r^2} \sigma_3\Big)^2 + \fft{dr^2}{f} +
\ft14 W r^2 \sigma_3^2 + \ft14 r^2 d\Omega_2^2\,,\qquad
A=\fft{\sqrt3\,q}{2r^2} \sigma_3\,,\nn\\
W &=& 1 - \fft{\alpha}{r^4}\,,\qquad f = (1+ g^2 r^2) W - \fft{g^2 q^2}{r^4}\,.
\eea
Mass and angular momentum are
\be
M=-\ft18\pi g^2 \alpha\,,\qquad J_\psi=\ft14\pi q\,.
\ee
The solution reduces to the static soliton (\ref{d5soliton1}) when $q=0$.  In order for the spacetime to avoid curvature singularity at $r=0$, there should be a Killing horizon at some $r_0>0$ such that $f(r_0)=0$.  Such a Killing horizon is guaranteed to exist if we have $\alpha>-g^2 q^2)$.  It follows that for given $q$, the mass of the solution has an upper bound, but no lower bound
\be
M<\ft18 \pi g^4 q^2\,.
\ee
This upper bound is analogous to (\ref{massup}).

It is clear that at the Killing horizon at $f=0$, we must have $W>0$.  It follows that there must be naked CTCs since
\be
g_{\psi\psi} = \fft{1}{4g^2} (f - W) = \ft14 \Big(r^2 W - \fft{q^2}{r^4}\Big)\,.
\ee
The manifold closes off at the Killing horizon provided that the null Killing vector on the horizon
\be
\ell=\fft{\sqrt{1 + g^2 r_0^2}}{2r_0^2 (1 + g^2 r_0^2)^2 + g^4 q^2}
\Big(q \fft{\partial}{\partial t} + 2r_0^2(1 + g^2 r_0^2) \fft{\partial}{\partial\psi}\Big)\,,
\ee
generates $2\pi$ period.  Note that in this time machine, then mass can be both positive and negative.  The electric charge vanishes, but there is a magnetic dipole charge on the Killing horizon
\be
{\cal D} = \fft{1}{8} \int_{r_0} F = \fft{\sqrt 3\,\pi\,q}{4r_0^2}\,.\label{dipole2}
\ee

\subsection{Asymptotic to flat spacetime}

We can turn off the cosmological constant and the solution becomes
\be
ds^2 =- \Big(dt - \fft{q}{2r^2} \sigma_3\Big)^2 + \fft{dr^2}{W} +
\ft14 W r^2 \sigma_3^2 + \ft14 r^2 d\Omega_2^2\,,\qquad
A=-\fft{\sqrt3\,q}{2r^2} \sigma_3\,.
\ee
This is a solution to field equations of five-dimensional minimal supergravity. The solution has zero mass but non-vanishing angular momentum
\be
M=0\,,\qquad J=\ft14\pi q\,.
\ee
The dipole charge takes the same form as (\ref{dipole2}). The metric describes a constant time bundle over the EH instanton, where the null Killing vector at $r=r_0=\alpha^{\fft14}$, namely
\be
\ell=\fft{q}{2r_0^2} \fft{\partial}{\partial t} + \fft{\partial}{\partial\psi}\,,
\ee
must generate $2\pi$ period.  Thus the spatial section is not asymptotic $\mathbb R^4$, but $\mathbb R^4/\mathbb Z_2$.

\section{Conclusions}

In this paper, we studied the properties of Kerr and Kerr-AdS metrics in $D=2n+1$ dimensions when they do not describe rotating black holes.  We found that when the mass was negative and all angular momenta turned on, the metrics could describe smooth time machines where spacetime closes off on some Euclidean pseudo horizon, which is Minkowski signatured, a time bundle over some base space. The absence of conical singularity of the degenerate surface of the horizon requires the periodic identification of the real time coordinate.  Such negative-mass time machines can arise for both asymptotically-flat or AdS spacetimes. We also constructed analogous time machines in gauged and ungauged minimal supergravity in five dimensions, where the time machines carry no electric but dipole charges.

Turning off the angular momenta appropriately, the aforementioned AdS time machines reduce to static solitons with negative mass. Furthermore, Euclideanization of Kerr-AdS metrics in odd dimensions can also lead to solitons with negative mass.  For those that are solutions to Einstein's vacuum field equations with or without a cosmological constant, the absence of any singularity implies that the origin of the spacetime curvature is purely gravitational without any matter energy-momentum tensor. This is very different from Schwarzschild or Kerr black holes where singular matter source is located at the singularity.  Thus our solutions are the manifestations of pure-gravity states.  Such states are not unusual in Euclidean signatured gravity; they are described by gravitational instantons.  Our work demonstrates that pure gravitational states can arise in Minkowski signatured gravity in $D=2n+1$. In addition, we find that taking the cosmological constant to zero, the AdS solitons solutions reduce to a class of Ricci-flat K\"ahler metrics in $D=2n$ dimensions.

Time machines are not unusual in supergravities where BPS time machines have been constructed.  What is unusual is perhaps that all these solutions carry negative energies.  It is thus of interest to examine the positive-mass conjecture.  Having naked CTCs can be perfectly consistent with the energy conditions.  In fact the naked CTCs in G\"odel-type metrics \cite{Godel:1949ga} emerge precisely because of the null-energy condition \cite{Li:2016nnn}.

Positive-mass conjecture states that mass of asymptotically Minkowski spacetime is non-negative. In our time-machine solutions, the time is required to be periodic. Although the asymptotic spacetime is flat for the $\Lambda=0$ solutions, it is not quite Minkowski, where time is isomorphic to a real line.  For the AdS solitons with negative mass, the EH instaton-like requirement of the period of $\psi$ coordinate implies that the asymptotic spacetime is AdS$/\mathbb Z_k$ rather than AdS.

Concrete examples of violating the positive-mass conjecture are perhaps those negative-mass AdS time machines. This is because in the flat-spacetime embedding of AdS, time in global coordinates are already periodic.  The further periodic identification of the null Killing vector on the Euclidean pseudo horizon can be perfectly consistent with the time period of global AdS provided that the constraint (\ref{cons1}) is satisfied. This implies that the mass and angular momenta are discretized and are functions of rational numbers. This phenomenon is analogous to the discretization of compact manifolds.  It can be argued that in the ``real world setting,'' spacetime configurations with discretized mass and angular momentum are so fine tuned and hence it is unlikely for the time machine to be created.  Of course, one can hardly call the AdS$_{2n+1}$ spacetime as the real world.  On the other hand, the discrete nature of the time-machine configurations suggests topological structures that imply that these solutions, although having negative mass, are stable.\footnote{We are grateful to Yi Wang for pointing this out.} It is of great interest to investigate the corresponding states in the boundary conformal field theory.

\section*{Acknowledgement}

We are grateful to Jianxin Lu, Chris Pope, Yi Wang, Zhao-Long Wang and Yu-Liang Wu for useful discussions. The work is supported in part by NSFC grants NO. 11475024, NO. 11175269 and NO. 11235003.

\end{document}